# Physical properties of Centaur (60558) 174P/Echeclus from stellar occultations

C. L. Pereira,[1,2]* F. Braga-Ribas,[1,2,3] B. Sicardy,[4] A. R. Gomes-Júnior,[2,5,6] J. L. Ortiz,[7]
H. C. Branco,[3,8] J. I. B. Camargo,[1,2] B. E. Morgado,[1,2,9] R. Vieira-Martins,[1,2] M. Assafin,[2,9]
G. Benedetti-Rossi,[2,6] J. Desmars,[10,11] M. Emilio,[1,3,12] R. Morales,[7] F. L. Rommel,[2,3] T. Hayamizu,[13,14]
T. Gondou,[13] E. Jehin,[15] R. A. Artola,[16] A. Asai,[13,14] C. Colazo,[17] E. Ducrot,[18] R. Duffard,[7] J. Fabrega,[19]
E. Fernandez-Valenzuela,[7,20] M. Gillon,[21] T. Horaguchi,[13,14,22] M. Ida,[13] K. Kitazaki,[13] L. A. Mammana,[23]
A. Maury,[24] M. Melita,[25,26,27] N. Morales,[7] C. Moya-Sierralta,[28] M. Owada,[13,14] J. Pollock,[29]
J. L. Sanchez,[30] P. Santos-Sanz,[7] N. Sasanuma,[13] D. Sebastian,[31] A. Triaud,[31] S. Uchiyama,[13] L. Vanzi,[32]
H. Watanabe[13] and H. Yamamura[13]

*Affiliations are listed at the end of the paper*

## ABSTRACT

The Centaur (60558) Echeclus was discovered on 2000 March 03, orbiting between the orbits of Jupiter and Uranus. After exhibiting frequent outbursts, it also received a comet designation, 174P. If the ejected material can be a source of debris to form additional structures, studying the surroundings of an active body like Echeclus can provide clues about the formation scenarios of rings, jets, or dusty shells around small bodies. Stellar occultation is a handy technique for this kind of investigation, as it can, from Earth-based observations, detect small structures with low opacity around these objects. Stellar occultation by Echeclus was predicted and observed in 2019, 2020, and 2021. We obtain upper detection limits of rings with widths larger than 0.5 km and optical depth of $\tau = 0.02$. These values are smaller than those of Chariklo's main ring; in other words, a Chariklo-like ring would have been detected. The occultation observed in 2020 provided two positive chords used to derive the triaxial dimensions of Echeclus based on a 3D model and pole orientation available in the literature. We obtained $a = 37.0 \pm 0.6$ km, $b = 28.4 \pm 0.5$ km, and $c = 24.9 \pm 0.4$ km, resulting in an area-equivalent radius of $30.0 \pm 0.5$ km. Using the projected limb at the occultation epoch and the available absolute magnitude ($H_v = 9.971 \pm 0.031$), we calculate an albedo of $p_v = 0.050 \pm 0.003$. Constraints on the object's density and internal friction are also proposed.

## 1 INTRODUCTION

The Centaurs are bodies with chaotic orbits that have perihelion and the semimajor axis between the orbits of Jupiter and Neptune (5.2 au $< q <$ 30 au and 5.2 au $< a <$ 30 au) and not being in 1:1 mean-motion resonance with any planet (Jewitt 2009). These bodies are believed to be the primary source of Jupiter-family comets (JFC). Approximately 1/3 of Centaurs are temporarily trapped in orbits close to the Sun, and 2/3 are expected to be expelled from the Solar System (Tiscareno & Malhotra 2003). This population of small bodies is notable due to the presence of cometary activity, which has been observed in about 13 per cent of Centaurs (Bauer et al. 2008). Furthermore, the perihelion distance of the active Centaurs is systematically smaller than that of the inactive ones, leading us to believe that thermal processes may be responsible for triggering mass loss mechanisms (Jewitt 2009).

The first Centaur object discovered was (2060) Chiron in 1977. With about 210 km in diameter, it is the second largest of its class (Lellouch et al. 2017). Chiron showed cometary activity in 1988, with coma and cometary tail detection in the following years. With this behaviour, it was designated as 95P/Chiron. Stellar occultations allowed the detection of structures analogous to collimated cometary jets (Ruprecht et al. 2015; Sickafoose et al. 2020) or rings, a hypothesis still being discussed (Ortiz et al. 2015; Braga-Ribas et al. 2023; Ortiz et al. 2023). (10199) Chariklo is the largest known Centaur, with about 250 km in diameter (Morgado et al. 2021). This object was the first small body in the Solar System where rings were confirmed (Braga-Ribas et al. 2013), followed by the dwarf planet Haumea (Ortiz et al. 2017) and the trans-Neptunian object (Morgado et al. 2023; Pereira et al. 2023). Among the Centaurs, the object 29P/Schwassmann-Wachmann 1 (SW1) is notable. Having a quasi-circular orbit with a semimajor axis of 5.98 au, it presents

*E-mail: chrystianpereira@on.br

**Table 1.** Orbital parameters and physical properties of 174P/Echeclus.

| $a$ (au) | $q$ (au) | $e$ | $i$ (deg) | $H$v | $P$ (h) | $D$ (km) | $p_v$ |
|---|---|---|---|---|---|---|---|
| 10.744 | 5.835 | 0.457 | 4.344 | $9.971 \pm 0.031$ | $26.785178 \pm 0.000001$ | $64.6 \pm 1.6$ | $0.052^{+0.0070}_{-0.0071}$ |

*Note.* Orbital elements of 174P/Echeclus from Jet Propulsion Laboratory (JPL) Small-Body Data base Browser JPL#109: a, semimajor axis; q, perihelion distance; e, eccentricity; i, inclination; Absolute magnitude $H$, calculated from Gaia DR3 Solar System objects (SSO; Tanga et al. 2023) using Morales et al. (2022) method. Diameter $D$ and albedo $p_v$ from Duffard et al. (2014); Rotational period $P$ from Rousselot et al. (2021).

frequent outbursts and high rates of carbon monoxide production (Senay & Jewitt 1994; Wierzchos & Womack 2020). Furthermore, SW1 is located in what we refer to as the 'Gateway', a transitional region between Centaur objects and JFC (Sarid et al. 2019).

The Centaur (60558) Echeclus, previously designated 2000 EC$_{98}$, was discovered by Terry Bressi of Spacewatch on 2000 March 03. From thermal observations using *Spitzer Space Telescope*, Stansberry et al. (2007) calculated a diameter of $83.6 \pm 15$ km and geometric albedo of $0.0383^{+0.0189}_{-0.0108}$. Bauer et al. (2013) derived an equivalent diameter of $59 \pm 4$ km and albedo $0.08 \pm 0.02$ from thermal infrared observations using Wide-field Infrared Survey Explorer. Duffard et al. (2014) obtained with PACS instrument on Herschel Space Observatory an equivalent diameter of $64.6 \pm 1.6$ km with an albedo of $0.052^{+0.007}_{-0.0071}$. Table 1 shows some of its orbital and physical properties.

Echeclus presented some cometary activity over the last few years with different intensities. The main outburst occurred between November and December 2005 at 13 au, when its brightness increased by ∼7 magnitudes (Choi, Weissman & Polishook 2006a). Besides being the major activity recorded so far in terms of brightness, the source of the mass ejection appears to be displaced from Echeclus itself, being at ∼ 55 000 km from the main body. This behaviour of the coma was attributed to the detachment of a fragment during the outburst, with this fragment being the primary source (Weissman et al. 2006; Choi et al. 2006b; Bauer et al. 2008; Fernández 2009). On the other hand, the decentralized coma could result from the ejection of material in three different regions and in different occasions: two short events followed by a third longer-lasting ejection (Rousselot et al. 2016).

On 2011 May 30, Jaeger et al. (2011) reported a new outburst while Echeclus approached the perihelion at 7.5 au, with a coma extending for one arc-minute (or ∼ 327 000 km). Subsequent observations taken in June 2011 showed a 40 arcsec (∼ 218 000 km) coma and a jet-like feature 6 arcsec (∼ 33 000 km) long. During the interval of these observations, a ∼3 visual magnitude brightening was reported. The outbursts reported in August 2016 presented a brightening amplitude of ∼3 magnitudes, without notable features in the coma (Miles 2016). After its perihelion in April 2015, Echeclus had the second largest outburst in brightness, with an amplitude of ∼4 magnitudes (James 2018). Observations using Faulkes North & South Telescopes and the NASA Infrared Telescope Facility just after the 2017 December outburst revealed a coma morphology asymmetric in the North–South direction that confirmed previous ideas about the dust properties (Kareta et al. 2019). Subsequent studies suggest that seasonal effects can cause material ejections. Also, the variegation of the colour index as a function of radial distance from the nucleus indicates that particles may have different sizes or compositions (Rousselot et al. 2021).

This paper presents the results of a double-chord stellar occultation observed in January 2020, a single-chord detection in January 2021, and an August 2019 close appulse. The 2020 event allowed constraining the size and shape of Echeclus. It was also used with the 3D model and pole coordinates position proposed by Rousselot et al. (2021) to obtain its size and volume. All data sets of the three stellar occultation events were used to search for signatures that could reveal the presence of secondary structures around Echeclus, whether coma, jets, arcs, or confined rings.

## 2 OBSERVATIONS

The stellar occultations presented in this work were predicted by the Lucky Star project[1] using the Numerical Integration of the Motion of an Asteroid (NIMA) ephemeris (Desmars et al. 2015) and *Gaia* Data Release 2 (*Gaia* DR2) sources (Gaia Collaboration 2018). The event analysis was made using the *Gaia* Data Release 3 (*Gaia* DR3) catalogue (Gaia Collaboration 2022). The geocentric position of the target star propagated from *Gaia* DR3 catalogue to the occultation epochs is presented in Table 2, and the observational circumstances of all observatories are presented in Table B1. When available, the data sets were calibrated for bias, dark, and flat fields using the Image Reduction and Analysis Facility (IRAF) (Butcher & Stevens 1981). The light curves were obtained through differential aperture photometry using the Platform for Reduction of Astronomical Images Automatically (PRAIA) package (Assafin et al. 2011).

Finally, the ingress and egress instants were obtained from a pipeline built using the Stellar Occultation Reduction Analysis library (SORA)[2] (Gomes-Júnior et al. 2022) by modelling the light curve considering a sharp-edge model convolved with Fresnel diffraction, the apparent star diameter at Echeclus distance, and integration time (more details in Braga-Ribas et al. 2013; Souami et al. 2020).

### 2.1 The Echeclus appulse on 2019 October 29

The event was predicted to cross the northern region of Chile, Argentina, Paraguay, and the southern part of Brazil. The geocentric closest approach occurred on 2019 October 29, at 08:40:50.2 UT, with a shadow velocity of 12.8 km s$^{-1}$. The maximum expected duration for the occultation was 6.8 s.

Observations for this event were conducted at Southern Astrophysical Research Telescope (SOAR, Cerro Pachón, Chile), TRAnsiting Planets and PlanetesImals Small Telescope (TRAPPIST-South, La Silla Observatory, Chile; Jehin et al. 2011), Search for Planets ECLipsing ULtra-cOOl Stars Io telescope (SPECULOOS-South Observatory, Cerro Paranal, Chile; Delrez et al. 2018; Jehin et al. 2018), New Technology Telescope (NTT, La Silla Observatory, Chile), and San Pedro de Atacama Celestial Explorations Observatory. Unfortunately, the main body shadow did not cross either of these sites, as shown in Fig. D1, and thus, the occultation by the Echeclus main body could not be detected. With the positions obtained from the stellar occultations of 2020 and 2021 and considering that the uncertainty

---
[1] https://lesia.obspm.fr/lucky-star/
[2] https://sora.readthedocs.io/

3626  *C. L. Pereira et al.***Table 2.** Stellar parameters from *Gaia* DR3 catalogue for each observed event. The equatorial coordinates were propagated to the event epoch.

| Epoch (UT) | *Gaia* DR3 Source ID | RA | Dec. | G (mag) |
| --- | --- | --- | --- | --- |
| 2019-10-29 08:40:50.1 | 3 395 392 378 744 301 696 | $05^h16^m34.39005^s \pm 0.0614$ mas | $+18°22'21.3283'' \pm 0.0746$ mas | 14.2 |
| 2020-01-22 01:44:51.8 | 3 406 534 520 342 323 584 | $04^h51^m35.92194^s \pm 0.1405$ mas | $+17°48'18.7786'' \pm 0.1481$ mas | 15.3 |
| 2021-01-19 09:10:44.6 | 3 399 123 330 937 092 224 | $05^h43^m05.01044^s \pm 0.0609$ mas | $+18°40'37.0924'' \pm 0.0897$ mas | 11.0 |

**Table 3.** Ingress and egress instants obtained from light curve modelling for 2020 January 22 and 2021 January 19 events. The $1\sigma$ error bars between parenthesis are in seconds.

| Event | Site | Ingress (UT) (hh:mm:ss.s) | Egress (UT) (hh:mm:ss.s) |
| --- | --- | --- | --- |
| 2020 | SOAR | 01:44:08.33 (0.01) | 01:44:10.15 (0.01) |
| 2020 | La Canelilla | 01:44:07.03 (0.04) | 01:44:11.80 (0.04) |
| 2021 | Anan | 09:15:03.173 (0.006) | 09:15:06.378 (0.009) |

**Table 4.** Data used in the equilibrium analysis.

| $\alpha(c/a)$ | $\beta(b/a)$ | Period (h) | $\phi$ (°) | $\rho$ (kg m$^{-3}$) |
| --- | --- | --- | --- | --- |
| $0.67 \pm 0.02$ | 0.77 | 26.785178 | 9 | 500–1,210 |
| | | | 10 | 600 |
| | | | 8.5–10.5 | 500–1,900 |

in the stellar position is very small (RUWE[3] = 0.948), we were able to reconstruct the geometry of the appulse of 2019. This allows us to determine the distance between the chords and the object's centre, establishing the detection limits for additional material. Fig. F1 displays the reconstructed map showing the theoretical path of the shadow based on the updated NIMAv9 ephemeris.

### 2.2 Stellar occultation on 2020 January 22

The 2020 January 22 stellar occultation was predicted to cross central Chile, Argentina, and Uruguay. The shadows velocity was 11.1 km s$^{-1}$, resulting in a maximum duration of 7.8 s. An observation campaign was triggered over South America involving professional and amateur observatories. Two out of nine observations were positive, with four being negative and three presenting technical problems/overcast. Thus, this was the first double-chord occultation for Echeclus.

The data acquisition with SOAR was made using the visitor instrument Raptor 247 Merlin camera with GPS as a time source. In La Canelilla/Chile, the data was acquired using a 520 millimeters telescope equipped with a ZWO ASI1600MM CMOS camera in the avi video format and NTP monitor as reference time. The video was converted to fits files using Python routines based on ASTROPY (Astropy Collaboration 2013). With the Fresnel scale $L_f = \sqrt{\lambda D/2} = 0.7$ km and the apparent star diameter at 8.66 au of about 0.1 km using the models provided by van Belle (1999), the light curves are dominated by the instrumental response, considering the exposure times of 0.25 s ($\sim 2.8$ km) for SOAR and 0.3 s ($\sim 3.3$ km) for La Canelilla/Chile. The ingress and egress instants obtained from modeling are shown in Table 3, and the modelled light curves are presented in Fig. E1.

### 2.3 Stellar occultation on 2021 January 19

The shadow's path for the stellar occultation on 2021 January 19 was predicted to pass over Japan. The star's brightness allowed small aperture telescopes to participate in this observational campaign. Unfortunately, the object's shadow path shifted with respect to the predicted path for about a radius to the south, reaching the $1\sigma$ uncertainty limit of the prediction. As a result, most of the telescopes involved in the observational campaign were outside the shadow-path region. The event was successfully observed through thin clouds at Anan Science Center in Anan, Tokushima, Japan. The acquisition was performed in video format using a ZWO ASI290MM camera coupled with a 254-mm telescope. An NTP monitor provided a reference time. For this event, the Fresnel Scale is calculated as $L_f = \sqrt{\lambda D/2} = 0.7$ km, and the star's apparent diameter was estimated to be 0.4 km at Echeclus distance using the models provided by van Belle (1999). With an exposure time of 0.028 s, the instrumental response is approximately 0.5 km. The instants of ingress and egress are presented in Table 3.

## 3 SIZE AND SHAPE

By analysing 27 light curves obtained between 2001 and 2019, Rousselot et al. (2021) reproduced a 3D model for Echeclus using the Shaping Asteroid models using Genetic Evolution (SAGE) modelling technique (Bartczak & Dudziński 2018), an algorithm based on light-curve inversion. The sidereal rotational period for Echeclus was determined to be $P = 26.785178 \pm 10^{-6}$ h. The rotational period error bar gives reasonable confidence in the orientation of the 3D shape at the time of occultation. The double-peak light curve exhibited amplitudes consistent with a triaxial body ($a > b > c$, where $c$ is the rotation axis), with a semimajor axial ratio of $a/b = 1.32$ and $b/c \sim 1.1$ based on a comparison of the projected areas of the 3D model (Rousselot et al. 2021).

The 3D model of Echeclus can be accessed through the interactive service for asteroid models (ISAM).[4] The rotational elements of the model are $\lambda = 115.2°$, $\beta = 21.5°$, and $\gamma_0 = 80°$ for the reference epoch $t_{ref} = 2455437.367$ JD, based on Kaasalainen, Torppa & Muinonen (2001) system. To fit the occultation chords to the 3D shape model, we used SORA. To orientate the 3D shape model for the occultation epoch, SORA uses the International Astronomical Union (IAU) recommendations by Archinal et al. (2018). We transformed the ISAM parameters to the IAU formalism using appropriate rotation matrices. In the IAU formalism, pole coordinates are RA = $122°$ $18'$ $05.8''$, Dec. = $42°$ $09'$ $16.6''$, and

---

[3]Renormalized Unit Weight Error (RUWE) indicates the reliability of the single-star model based on observations, with values close to 1 being expected. Values greater than 1.4 may suggest that the source is not a single star or that there are issues with the astrometric solution (Lindegren et al. 2018).

[4]http://isam.astro.amu.edu.pl/

MNRAS **527**, 3624–3638 (2024)

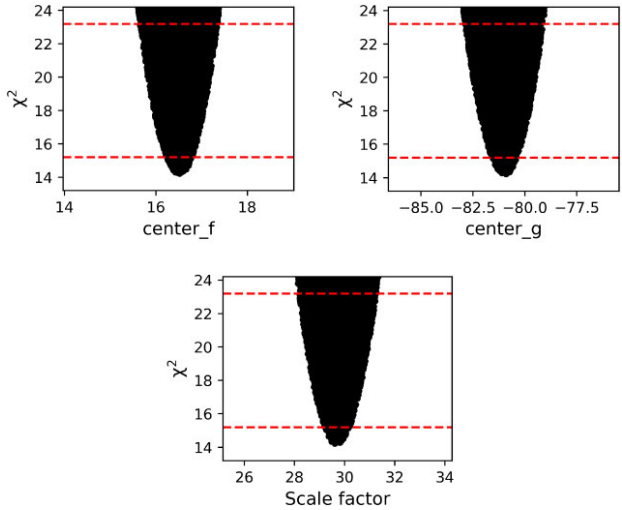

**Figure 1.** $\chi^2$ distributions obtained from shape fit for the centre positions (f and g) and scaling factor for 2021 stellar occultation. The red horizontal dashed lines delimits the $1\sigma$ (lower line) and $3\sigma$ (upper line) uncertainties.

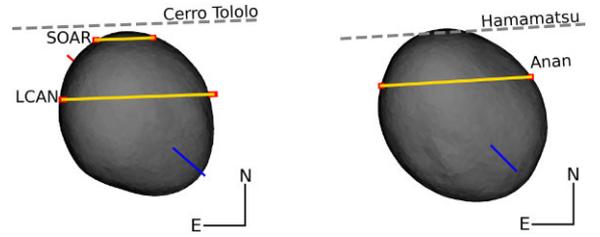

**Figure 2.** Plane-of-sky views of Echeclus 3D model with the observed chords, being the 2020 observed chords on the left and the 2021 observed chord on the right.

$W_0 = 336.793618$ degrees in the reference epoch $t_{ref} = 2455437.367$ JD.

Once the apparent shape of Echeclus projected onto the sky plane at the occultation epoch is obtained, the positive chords of the 2020 event can be compared with the projected limb. This comparison involves minimizing the $\chi^2$ function defined by

$$\chi^2 = \sum_{i=1}^{N} \frac{(\phi_{i,obs}) - \phi_{(i,cal)}}{\sigma_i^2}, \quad (1)$$

where $\phi_{(i,obs)} - \phi_{(i,cal)}$ is the difference between the extremity of each chord and the projected shape limb and $\sigma_i$ is the uncertainty of the $i$-th chord extremity. We generate numerous models by applying an offset in the body centre and re-scaling the model iteratively. This re-scaling procedure is necessary because some objects do not have a known radius, and SAGE usually scales the shape model so that the maximum radius equals 1. The obtained values are $f_c = 16.5 \pm 0.3$ km, $g_c = -80.7 \pm 0.6$ km, and a scale factor of $29.4 \pm 0.5$, with the correspondent $\chi^2$ curves presented in Fig. 1. Note that there are differences between the model and the chords, but this can be accommodated by reasonable topographic features not considered in the 3D model. Checking the radial residuals between the extremities of the 2020 occultation chords and the limb, we obtain 0.3 and 0.6 km (respectively, 1.1 and 1.0 km) for the SOAR chord (respectively La Canelilla), from West to East. This is about 2.6 per cent of the estimated equivalent radius for the 3D model.

A similar procedure was applied to compare the positive chord of the 2021 stellar occultation with the proposed 3D model and pole. In this case, the better correspondence between chord extremities and projected limb is obtained by varying the center position while keeping the same scale obtained from the 2020 occultation. The results are presented in Fig. 2.

From the obtained scale, we can calculate the dimension of Echeclus' axes, being as $a = 37.0 \pm 0.6$ km, $b = 28.4 \pm 0.5$ km, and $c = 24.9 \pm 0.4$ km, giving an area-equivalent radius of $R_{equiv} = 30.0 \pm 0.5$ km. The geometric albedo was calculated using $p_v = 10^{0.4(H_{\odot,v}-H_v)} \cdot (au_{km}/R_{equiv})^2$ (Sicardy et al. 2011), where $au_{km} = 1$ au $= 1.49598 \times 10^8$ km, $H_{\odot,v} = -26.74$ is the Sun absolute magnitude in $v$ band, $H_v$ is the Echeclus instantaneous absolute magnitude in $v$ band at the rotational phase, and $R_{equiv} = 30.0 \pm 0.5$ km is the area-equivalent radius of projected Echeclus' limb at the occultation epoch. The absolute magnitude $H_v = 9.971 \pm 0.31$ is based in 438 multi-epoch observations from *Gaia* SSO release (Tanga et al. 2023) and was calculated using the tools and procedures described in Morales et al. (2022). Considering the rotational light curve and the rotational period published by Rousselot et al. (2021), we were able to determine the rotational phase and, therefore, the geometric albedo for Echeclus at 2020 stellar occultation instant as $p_v = 0.050 \pm 0.003$.

## 4 EQUILIBRIUM SHAPE AND DENSITY ANALYSIS

The equilibrium shape analysis method proposed by Holsapple (2001, 2004, 2007) can investigate small bodies' physical properties and internal structure of granular media characterized by an angle of friction. It correlates the object's dimensions (semi-axis $a$, $b$, and $c$), its angle of internal friction ($\phi$), and the scaled spin $\Omega$, defined by:

$$\Omega = \sqrt{\frac{4\pi}{P^2 \rho G}}, \quad (2)$$

where $\rho$ is the mean density in kg m$^{-3}$, $P$ the rotational period in hours, and $G$ the gravitational constant. It results in rough estimates for objects with known global composition and unknown density. For objects with unknown global composition and known density, it constrains the global composition and internal structure. The angle of internal friction measures the material's response to shear stress. Higher values of $\phi$ imply a higher resistance to shear stress and, hence, to deformation. Materials with fluid-like behaviour have $\phi$ of $0°$, easily deforming, while rocks often have $\phi > 20°$. For small bodies, it can be assumed based on the object's surface composition (when known) and/or the expected characteristics of its family.

Echeclus' composition is poorly known, resulting in very loose constraints of its $\phi$. An assumed range of densities was used to investigate Echeclus interior properties to obtain meaningful results. The resulting range of possible $\phi$ was used to (i) base general considerations regarding the object's global composition and internal structure and (ii) to indicate average global density estimates for chosen $\phi$ values.

For a global density between 500 and 2500 kg m$^{-3}$, a reasonable range of densities for Centaurs and TNOs (Sicardy et al. 2011), Echeclus must have internal friction between $8.5° > \phi > 10.5°$ (Fig. 3), compatible with what is expected for a mixture of ice and rock (Barucci, Doressoundiram & Cruikshank 2004; Yasui & Arakawa 2009). For comparison, Ortiz et al. (2017) determined a friction angle for Haumea between $10°$ and $15°$, considering the c/a = 0.4 ratio. Homogeneous objects behave as ideal fluids, with $\phi$ close or equal to $0°$, while heterogeneous bodies don't (Holsapple 2001). Hence, it is likely that Echeclus has a heterogeneous internal structure. Given its small size and the consequent unlikeliness that it had sufficient internal heat to experience any significant degree of differentiation (Shchuko et al. 2014), it is possible that it has a heterogeneous distribution of mass, a rubble-pile structure or similar, with significant internal porosity may be as high as 20–30 per cent in total volume (Yasui & Arakawa 2009).

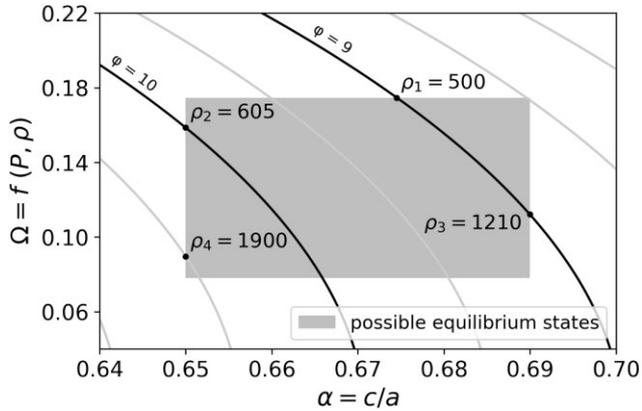

**Figure 3.** $\Omega$ versus $\alpha$ plot used for the equilibrium shape analysis. The grey lines correspond to different $\phi$, spaced every 0.5°. The black lines indicate the curves for $\phi = 9°$ and 10°, as labelled. The grey box, limited in the x-axis by the observed c/a from the scaled 3D shape, indicates the range of possible equilibrium states for Echeclus. Its vertical size is defined by variations in the scaled spin $\Omega$, a function of the object's period of rotation (P) and density ($\rho$). Letting P be constant, $\Omega$ varies with $\rho$. The black points indicate the corresponding $\rho$ (in kg m$^3$) for the combination of $\Omega$ and $\alpha$.

## 5 LIMITS ON ADDITIONAL MATERIAL AROUND ECHECLUS

Detection limits can be calculated considering that the standard deviation of the light curve in the regions external to central body occultation is related to the minimum detectable apparent opacity p$'$ of any structure around a body (Braga-Ribas et al. 2023, and references therein). The search for additional material was done using light curves from the three events here reported, to determine the detection limits of apparent opacity (p$'$) and optical depth ($\tau'$). We call these values *apparent* since the properties are measured in the sky plane and not perpendicular to the structures, for example, in the case of discs or rings.

The depth of the observed stellar flux drop is the apparent opacity and is defined by $p' = 1 - \frac{I}{I_0}$, where I and $I_0$ are the transmitted and incident stellar flux, respectively. Assuming that the white noise follows a Gaussian distribution and considering that the standard deviation of the target star flux is equivalent to p$'$, a single point detection will only be statistically significant if it is outside the 3$\sigma$ level of the mean flux. With the apparent opacity, we can constrain the apparent optical depth from $\tau' = -ln(1 - p')$ for a possible structure around Echeclus (more details in Boissel et al. 2014 and Bérard et al. 2017).

A disc is made of particles that diffract light from the star individually. For a particle in the ring with a radius of approximately 1 m, we have a diffraction cone with an angular diameter $\phi_d = \lambda D/r$ of about 900 km, at the Echeclus distance ($1.3 \times 10^9$ km) and for observations at a wavelength centered at 0.65 µm. Even for the 100 km smoothed light curve, the structure widths we seek are narrower than the diffraction cone-diameter. This implies a loss of energy both by the individual diffraction of the particles and by the flux blocked by these particles when these are much larger than the wavelength. This is known as the extinction paradox. Thus, the rings appear twice as optically thick as they are (Cuzzi 1985). In this sense, the real optical depth in the sky plane is $\tau = \tau'/2$, or equivalently for the apparent opacity, $p = 1 - \sqrt{1 - p'}$.

Data points on a light curve above the 3$\sigma$ level are analysed individually against the reference star flux to eliminate the possibility that these drops in flux could have been caused by passing clouds or other artefacts in the image. To search for wider structures, such as the broad, shallow drops in flux observed around Chiron in a 1993 stellar occultation Elliot et al. (1995), we resampled our best light curves for each event by applying a Savitzky–Golay filter with windows of ∼100 km (F3 extension from Elliot et al. 1995). The detection limit for all light curves in the three events analysed in this work is presented in Table C1.

We can also compute limits for apparent equivalent width, which will give us the minimum values for the length of a chord from a serendipitous stellar occultation and, therefore, a upper limit for the radius of the occulting body. Ignoring diffraction effects, we can obtain this limit from the equation $E'_p(i) = [1 - \phi(i)]\Delta r(i)$, where $\phi(i)$ is the normalized stellar flux and $\Delta r(i)$ is the difference between consecutive points in the sky plane, obtained from exposure time times sky plane radial velocity. The values for $E'_p$ are presented in Table C1.

The best data set from the 2019 appulse was obtained with NTT at La Silla Observatory, with an average radial resolution of 2.6 km and a standard deviation of 0.048 (1$\sigma$). We determine the 3$\sigma$ level for the optical depth as $\tau = 0.07$ and apparent equivalent width as $E'_p \sim 380$ m. For the smoothed light curve, we estimated limits on apparent optical depth as $\tau = 0.01$, covering about 7,123 km radially. As mentioned, we did not detect the main body at the observatories involved in this campaign; therefore, to determine the radial distances, we considered the object's theoretical position calculated with NIMA ephemeris, which used the astrometric positions obtained from the 2020 and 2021 detected occultations.

In the 2020 event, the best light curve was obtained with the SOAR telescope. This light curve is positive, so the standard deviation was calculated in the regions outside the main body occultation. The calculated limits for apparent optical depth and apparent equivalent width in the full resolution light curve (average 2.9 km per data point) is $\tau = 0.15$ and $E'_p \sim 800$ m. For the resampled light curve, with a spatial resolution of 97.8 km, the 3$\sigma$ limit for apparent optical depth is $\tau = 0.04$. The radial sky plane cover is about 14 000 km.

Since the star involved in the 2021 stellar occultation is bright, a high acquisition rate could be used on some telescopes. The best light curve considering S/N and spatial resolution was obtained at Okazaki/JP, with $\delta_r = 1.2$ km and standard deviation at 1$\sigma$ level of 0.10. The full-resolution light curve provides a lower limit for the apparent optical depth of $\tau = 0.16$. The light curve resampled over a 99.4 km window has a detection limit of $\tau = 0.02$ and $E'_p \sim 370$ m, with the chord passing distant 25 km from Echeclus center and covering a distance of 9600 km in the sky plane.

## 6 CONCLUSION AND DISCUSSION

Echeclus, a Centaur with cometary activity, has shown multiple outbursts and has been the subject of extensive observations. In this work, we presented the results of the two successful stellar occultations events by Echeclus in 2020 and 2021, and an appulse in 2019. These observations and recent publications with photometric studies allow us to obtain important physical properties of this Centaur. In addition, better-quality light curves made it possible to probe the surroundings of the object in search of confined or diffuse structures.

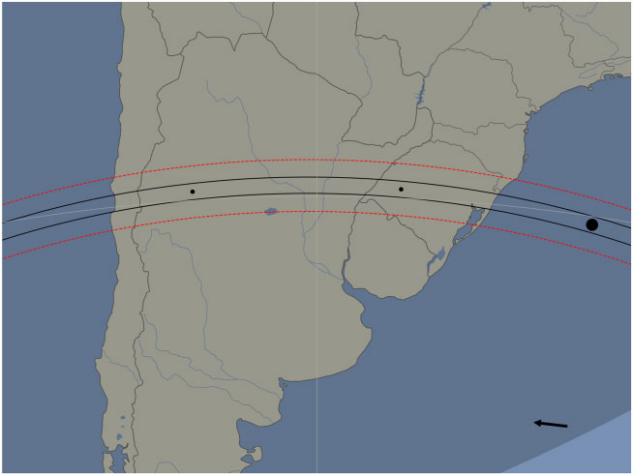

**Figure 4.** Echeclus stellar occultation predicted for 2023 December 09, at 05:09 UT. This event involves a star with a magnitude G = 15.6. More details on this occultation can be found in Lucky Star page. The black lines indicate the Echeclus shadow limits and the red dashed lines indicate the $1\sigma$ uncertainty along the shadow path. The black dots are spaced by one minute and the big black dot corresponds to the geocentric closest approach (2023-12-09 05:09:13). The arrow indicates the shadow direction.

We use the 3D model and pole orientation ($\lambda, \beta = 115.2°, 21.5°$) for Echeclus from ISAM service and its rotational period (P = $26.785178 \pm 10^{-6}$ h) proposed by Rousselot et al. (2021) to fit the projected limb to the stellar occultation chords. As the 2020 event has two positive chords, we propagate the rotation of Echeclus from the model's reference epoch to the epoch of the occultation. By varying the limb center and scale, we calculate the values for the semi-axes of Echeclus, being: $a = 37.0 \pm 0.6$ km, $b = 28.4 \pm 0.5$ km, and $c = 24.9 \pm 0.4$ km, giving us an area-equivalent radius of $30.0 \pm 0.5$ km. Using the area of the projected limb and the rotational phase at the occultation epoch, we determine the instantaneous geometric albedo as $p_v = 0.050 \pm 0.003$.

Assuming reasonable densities for Echeclus, we used the method proposed by Holsapple (2001, 2004, 2007), where the relationship between spin, shape, and assumption of certain geological properties was used to constrain the object's internal structure. Echeclus has a triaxial shape with an internal friction angle $8.5° > \phi > 10.5°$. The range of possible densities can be further constrained by assuming further limits for $\phi$ (Fig. 4). For $\phi = 9°$, Echeclus must have a density of $500 - 1,210$ kg m$^{-3}$. For $\phi = 10°$, it must have a density superior to $600$ kg m$^{-3}$. Lastly, $8.5° \phi < 10.5°$ must have a density between 500 and 1900 kg m$^{-3}$. These values indicate that Echeclus comprises ice and silicates, with a low proportion of ice and significant porosity, as rubble-pile or layered-pile (Belton et al. 2007) structures. Based on Rosenberg's study of comet 67P, Rousselot et al. (2021) propose that a layered structure and the internal inhomogeneity of Echeclus may be directly related to the erratic ejections of material observed by this object, with outbursts of different intensity and duration.

We searched for significant flux drops in the stellar occultation light curves, indicating sparsed or confined material around Echeclus. Here, we consider the standard deviation at the $3\sigma$ level to be the upper limit for the apparent opacity p' (on the sky plane), so we can estimate the limits for the actual optical depth $\tau$. In 2019, the best data set gave us an optical depth limit of $\tau = 0.07$, considering structures with a radial width of 2.6 km in the sky plane. In the case of opaque structures, small satellites with a diameter greater than 380 meters would be detectable in the regions probed by the chords.

In 2020, the best light curve allowed structures with a radial width 2.9 km in the sky plane to be detected if they have an optical depth $\tau > 0.15$, or opaque small satellites with a minimum diameter of 800 m. For 2021, structures with a radial width of 1.2 km in the sky plane would be detected with $\tau > 0.16$, or opaque structures with ∼370 m in diameter. These values are for the best data sets presented in this work. However, it is clear that the detection limits of opaque structures, such as small satellites, need to be considered individually. Each light curve sweeps a different region of the Echeclus neighbourhoods and is susceptible to detecting a small satellite.

During the detection limit procedure, we observed some points with flux drops greater than the $3\sigma$ standard deviation. These notable points were individually analysed against calibration light curves. This eliminates the possibility that these drops were caused by artefacts in the image that were not fully mitigated during the construction of the flux ratio light curve. From this analysis, we noted that these individual points outside $3\sigma$ on flux standard deviation are related to passing clouds or seeing degradation. So, if there is scattered material around Echeclus, it does not have enough optical depth to be detected in the light curves of stellar occultations obtained so far.

Finally, observations of new stellar occultations by Echeclus are needed so that the 3D model can be improved. Therefore, multichord stellar occultations are essential. Indeed, a promising event should occur on 2023 December 9, involving a star with magnitude G = 15.8, with the shadow passing over southern Brazil, northern Argentina, and La Serena region in Chile (Fig. 4).

**ACKNOWLEDGEMENTS**

CLP is thankful for the support of the Coordenação de Aperfeiçoamento de Pessoal de Nível Superior - Brasil (CAPES) and Fundação de Amparo à Pesquisa do Estado do Rio de Janeiro - FAPERJ/DSC-10 (E26/204.141/2022). This work was carried out within the 'Lucky Star' umbrella that agglomerates the efforts of the Paris, Granada, and Rio teams, funded by the European Research Council under the European Community's H2020 (ERC grant agreement no. 669416). This study was partly financed by the National Institute of Science and Technology of the e-Universe project (INCT do e-Universo, CNPq grant 465376/2014-2). This study was financed in part by CAPES – Finance Code 001. The authors acknowledge the respective CNPq grants: BEM 150612/2020-6; FBR 314772/2020-0; RVM 307368/2021-1; MA 427700/2018-3, 310683/2017-3, 473002/2013-2; JIBC acknowledges grants 305917/2019-6, 306691/2022-1 (CNPq), and 201.681/2019 (FAPERJ). JLO acknowledges financial support from the grant CEX2021-001131-S funded by MCIN/AEI/ 10.13039/501100011033, also acknowledges the financial support by the Spanish grants PID2020-112789GB-I00 from AEI and Proyecto de Excelencia de la Junta de Andalucía PY20-01309. PSS acknowledges financial support from the Spanish I+D+i project PID2022-139555NB-I00 funded by MCIN/AEI/10.13039/501100011033. We acknowledge financial support from the Severo Ochoa grant CEX2021-001131-S funded by MCIN/AEI/10.13039/501100011033. ED acknowledges support from the innovation and research Horizon 2020 programme in the context of the Marie Sklodowska-Curie subvention 945298. TRAPPIST-South is funded by the Belgian Fund for Scientific Research (Fond National de la Recherche Scientifique, FNRS) under the grant PDRT.0120.21. EJ is F.R.S.-FNRS Senior Research Associate. This research received funding from the European Research Council (ERC) under the European Union's Horizon 2020 research and innovation programme (grant agreement no. 803193/BEBOP), from the Science and Technology Facilities Council (STFC; grant no.

ST/S00193X/1), and from the MERAC foundation . The ULiege's contribution to SPECULOOS-South Observatory has received funding from the European Research Council under the European Union's Seventh Framework Programme (FP/2007–2013) (grant Agreement no. 336480/SPECULOOS), from the Balzan Prize and Francqui Foundations, from the Belgian Scientific Research Foundation (F.R.S.-FNRS; grant no. T.0109.20), from the University of Liege, and from the ARC grant for Concerted Research Actions financed by the Wallonia-Brussels Federation. The contribution of the University of Cambridge to SPECULOOS-South Observatory is supported by a grant from the Simons Foundation (PI Queloz, grant number 327127). MG is F.R.S.-FNRS Research Director LV was partially funded by ANID, BASAL, FB210003. We thank the observers A. Hashimoto, M. Meunier for their efforts in the observational campaigns. Some results were based on observations taken at the 1.6 m telescope at Pico dos Dias Observatory of the National Laboratory of Astrophysics (L.N.A./Brazil). This work has made use of data from the European Space Agency (E.S.A.) mission *Gaia* (https://www.cosmos.esa.int/gaia), processed by the *Gaia* Data Processing and Analysis Consortium (D.P.A.C., https://www.cosmos.esa.int/web/gaia/dpac/consortium).

## DATA AVAILABILITY

Data available on request.

## APPENDIX A: ASTROMETRIC POSITIONS

The astrometric positions obtained for Echeclus for the geocentric closest approach instant in each stellar occultation event are presented in A1.

**Table A1.** Astrometric positions of Echeclus based on the 3D model centre for the geocentric closest approach epoch.

| Epoch (UT) | RA | Dec. |
| --- | --- | --- |
| 2020-01-22 01:44:49.9 | $04^h\ 51^m\ 35^s.9201947 \pm 0.146$ mas | $17°\ 48^m\ 18^s.024343 \pm 0.171$ mas |
| 2021-01-19 09:10:44.3 | $05^h\ 43^m\ 05^s.0114595 \pm 0.063$ mas | $18°\ 40^m\ 37^s.397107 \pm 0.093$ mas |

# APPENDIX B: OBSERVATIONAL CIRCUMSTANCES

The observational circumstances for each site involved in the stellar occultation campaigns presented in this work are listed in Table B1.

In total, we have three positive light curves, fourteen negatives, and nine sites with no data provided due to technical problems or a cloudy sky.

**Table B1.** Observational circumstances for the occultation campaigns between 2019 and 2021. It presents the site name, geographic coordinates, telescope aperture, the detector configuration (filter, exposure time, and cycle), and the observer names.

| Site | Latitude (N) Longitude (E) Altitude (m) | Telescope aperture (m) Detector Filter | Exposure time Cycle (s) | Observers | $\sigma_{flux}$ Status |
|---|---|---|---|---|---|
| | | **2019 October 29 – South America** | | | |
| SOAR | −30°14′16.7″ | 4.1 | 0.4 | J. Camargo | 0.035 |
| Cerro Pachón | −70°44′01.0″ | Raptor/Merlin127 | 0.4 | A.R. Gomes-Jr | Negative |
| Chile | 2,738 | Clear | | F. Rommel | |
| TRAPPIST-South | −29° 15′16.6″ | 0.6 | 1.5 | E. Jehin | 0.020 |
| La Silla | −70° 44′21.8″ | FLI PL3041-BB | 2.2 | E. Ducrot | Negative |
| Chile | 2,337 | Clear | | | |
| SPECULOOS – South (SSO) | −24° 36′57.9″ | 1.0 | 1.0 | | 0.010 |
| Cerro Paranal | −70° 23′26.0″ | Andor Tech | 2.2 | E. Jehin | Negative |
| Chile | 2,479.2 | Clear | | | |
| Celestial Explorations Observatory | −22° 57′09.8″ | 0.4 | 3.0 | A. Maury | 0.035 |
| San Pedro de Atacama | −68° 10′48.7″ | FLI PL16803 | 4.7 | J. Fabrega | Negative |
| Chile | 2,396.9 | Clear | | | |
| NTT | −29° 15′32.1″ | 3.58 | 0.2 | | 0.045 |
| La Silla | −70° 44′01.5″ | SofI | 0.2 | B. Sicardy | Negative |
| Chile | 2,375 | H | | | |
| ASH2 | −22° 57′09.8″ | 0.4 | – | | |
| San Pedro de Atacama | −68° 10′48.7″ | STL 11000 | – | N. Morales | No data |
| Chile | 2,396.9 | Clear | | | |
| | | **2020 January 22 – South America** | | | |
| SOAR | −30°14′16.7″ | 4.1 | 0.25 | J. Camargo | 0.063 |
| Cerro Pachón | −70°44′01.0″ | Raptor/Merlin127 | 0.25 | A.R. Gomes-Jr | Positive |
| Chile | 2,738 | Clear | | F. Rommel | |
| La Canellila | −30° 32′01″ | 0.520 | 0.3 | | 0.152 |
| Limarí | −70° 47′46″ | ZWO ASI1600 MM | 0.3 | M. Meunier | Positive |
| Chile | 1,570 | Clear | | | |
| Geminis Austral Obs. | −32° 59′3.84″ | 0.4 | 2.0 | | 0.152 |
| Rosario | −60° 39′29.28″ | ZWO ASI1600 MM | 2.0 | J. L. Sanchez | Negative |
| Argentina | 22 | Clear | | | |
| CTIO | −30° 10′03.7″ | 1.0 | 0.5 | | 0.031 |
| Cerro Tololo | −70° 48′19.3″ | ? | 1.8 | J. Pollock | Negative |
| Chile | 2,286 | Clear | | | |
| PROMPT P5 | −30° 10′03.7″ | 0.4 | 0.5 | | 0.115 |
| Cerro Tololo | −70° 48′19.3″ | U47-MB | 1.2 | J. Pollock | Negative |
| Chile | 2,286 | Clear | | | |
| Santa Martina | −33° 16′09.0″ | 0.4 | 1.5 | L. Vanzi | 0.076 |
| Santiago | −70° 32′04.0″ | Raptor/Merlin127 | 1.5 | C. M. Sierralta | Negative |
| Chile | 1,450 | Clear | | | |
| Córdoba Ast. Obs. | −31° 25′12.2″ | 0.4 | – | | No data |
| Córdoba | −64° 11′55.1″ | QHY6 | – | R. Artola | Technical |
| Argentina | 427 | Clear | | | Problems |
| ASH2 | −22° 57′09.8″ | 0.4 | – | | |
| San Pedro de Atacama | −68° 10′48.7″ | STL 11000 | – | N. Morales | No data |
| Chile | 2,396.9 | Clear | | | |
| | | **Information about all sites** | | | |
| TRAPPIST-South | −29° 15′16.6″ | 0.6 | – | | No data |
| La Silla | −70° 44′21.8″ | FLI PL3041-BB | | E. Jehin | Overcast |
| Chile | 2,337 | Clear | | | |
| El Gato Gris Observatory | −31° 21′24.58″ | 0.35 | – | | No data |
| Tanti | −64° 35′34.41″ | QHY 174M | – | C. Colazo | Overcast |

**Table B1** – *continued*

| Site | Latitude (N) Longitude (E) Altitude (m) | Telescope aperture (m) Detector Filter | Exposure time Cycle (s) | Observers | $\sigma_{\text{flux}}$ Status |
|---|---|---|---|---|---|
| Argentina | 864 | Clear | | | |
| CASLEO (Jorge Sahade Telescope) | −31° 47′54.7″ | 2.15 | – | L. A. Mammana | No data |
| San Juan | −69° 17′44.1″ | Roper Versarray 2048B | – | M. Melita | Overcast |
| Argentina | 2,552.0 | Clear | | | |
| CASLEO (HSH Telescope) | −31° 47′54.7″ | 0.6 | – | L. A. Mammana | No data |
| San Juan | −69° 17′44.1″ | SBIG STL-1001E | – | M. Melita | Overcast |
| Argentina | 2,552.0 | Clear | | | |
| NTT | −29° 15′32.1″ | 3.58 | – | | No data |
| La Silla | −70° 44′01.5″ | SofI | – | None | Technical Problems/ |
| Chile | 2,375.0 | | | | Overcast |
| **2021 January 19 – Japan** | | | | | |
| Anan Science Center | 33°56′56.2″ | 0.254 | 0.0276 | | 0.225 |
| Tokushima | 134°40′23.0″ | ZWO ASI290MM | 0.0278 | T. Gondou | Positive |
| Japan | 15.3 | Clear | | | |
| Hamamatsu | 34°43′07.0″ | 0.250 | 0.269 | | 0.57 |
| Shizuoka | 137°44′23.0″ | WAT-120N + | 0.300 | M. Owada | Negative |
| Japan | 17 | Clear | | | |
| Toyohashi | 34°49′40.2″ | 0.2 | 0.246 | | 4.13 |
| Aichi | 137°26′09.0″ | ZWO ASI290MM | 0.246 | H. Yamamura | Negative |
| Japan | 8 | Clear | | | |
| Okazaki | 34°56′06.0″ | 0.2 | 0.071 | | 0.086 |
| Aichi | 137°08′33.2″ | ZWO ASI290MM | 0.071 | M. Ida | Negative |
| Japan | 20 | Clear | | | |
| Inabe | 35°10′14.7″ | 0.355 | 0.048 | | |
| Mie | 136°31′24.7″ | ZWO ASI290MM | 0.048 | A. Asai | Negative |
| Japan | 187 | Clear | | | |
| Inabe | 35°07′10.2″ | 0.130 | 0.33 | | |
| Mie | 136°33′33.9″ | WAT-120N | 0.33 | H. Watanabe | Negative |
| Japan | 92 | Clear | | | |
| Musashino | 35°42′36.9″ | 0.4 | 0.013 | | 0.156 |
| Tokyo | 139°33′41.3″ | ZWO ASI290MC | 0.013 | K. Kitazaki | Negative |
| Japan | 66 | Clear | | | |
| Kashiwa | 35°52′08.44″ | 0.250 | 0.033 | | 0.114 |
| Chiba | 139°58′50.67″ | WAT-910HX | 0.033 | S. Uchiyama | Negative |
| Japan | 20 | Clear | | | |
| Chichibu | 35°58′04.5″ | 0.4 | visual | | |
| Saitama | 139°01′59.6″ | No cam | visual | A. Hashimoto | Negative |
| Japan | 355 | No filter | | | |
| National Museum of Nature and Science | 36°06′05.1″ | 0.5 | 0.033 | | |
| Ibaraki | 140°06′40.7″ | WAT-910HX | 0.033 | T. Horaguchi | Negative |
| Japan | 40 | Clear | | | |
| Toshima-ku | 35°44′14.2″ | 0.3 | – | | |
| Tokyo | 139°44′35.8″ | WAT-910HX | – | N. Sasanuma | No data |
| Japan | 31 | Clear | | | |

## APPENDIX C: DETECTION LIMITS

Table C1 presents the results from detection limits determination calculated as explained in Section 5. W stands for width in the sky plane ($\delta_r$ for original resolution light curves), $\tau = \tau'/2$ is the optical depth, obtained from $\tau' = -\ln(1 - p'(3\sigma))$, and $E'_p$ is the apparent equivalent width. The sky cover is not necessarily centered in the body's position.

**Table C1.** Detection limits calculated for all light curves provided in this work for original spatial resolution and resampled data.

| Date | Site | Original resolution | | | Re-sampled data | | Sky cover (km) |
|---|---|---|---|---|---|---|---|
| | | $\delta_r$ (km) | $\tau$ | $E'_p$ (km) | W (km) | $\tau$ | |
| 2019-10-29 | SOAR | 5.3 | 0.06 | 0.65 | 94.8 | 0.02 | 12,397 |
| | San Pedro de Atacama | 39.6 | 0.06 | 4.47 | 79.1 | 0.04 | 30,970 |
| | TRAPPIST-South | 19.8 | 0.03 | 1.28 | 98.8 | 0.02 | 9,385 |
| | SPECULOOS-South Obs. (SSO) – Io | 13.2 | 0.02 | 0.44 | 92.3 | 0.01 | 9,851 |
| | NTT | 2.6 | 0.07 | 0.38 | 97.5 | 0.01 | 7,122 |
| 2020-01-22 | CTIO | 5.7 | 0.05 | 0.59 | 92.8 | 0.01 | 14,210 |
| | PROMPT-P5 | 5.7 | 0.21 | 2.21 | 97.8 | 0.05 | 9,930 |
| | SOAR | 2.9 | 0.15 | 0.80 | 97.8 | 0.04 | 13,809 |
| | Santa Martina | 17.2 | 0.13 | 4.38 | 86.2 | 0.06 | 13,782 |
| | La Canelilla | 3.4 | 0.28 | 1.79 | 96.7 | 0.06 | 1,723 |
| 2021-01-19 | Anan Sci. Center | 0.5 | 0.39 | 0.32 | 99.7 | 0.02 | 576 |
| | Hamamatsu | 2.6 | 0.11 | 0.54 | 96.5 | 0.02 | 2,031 |
| | Toyohashi | 4.2 | 0.08 | 0.69 | 96.5 | 0.02 | 9,207 |
| | Okazaki | 1.2 | 0.16 | 0.37 | 99.4 | 0.02 | 9,614 |
| | Kashiwa | 0.6 | 0.24 | 0.24 | 99.9 | 0.09 | 3,093 |
| | Musashino | 2.2 | 0.14 | 0.59 | 99.0 | 0.02 | 14,359 |
| | Inabe | 4.6 | 0.14 | 1.24 | 96.0 | 0.04 | 3,498 |
| | National Museum of Nat. and Sci. | 0.6 | 0.16 | 0.17 | 99.4 | 0.02 | 3,718 |

## APPENDIX D: LIGHT CURVES

We present the light curves of the radial distance from Echeclus center versus flux (Figs D1, D2, and D3) with all data sets provided by observers, be positive or negative detections. These light curves are organized in these plots from north to south. Data recording at some sites was interrupted just before the local closest approach, for this reason, they are not shown in this graph. Also, we remove the data points within the ingress and egress interval to better view the flux standard deviation at $3\sigma$ level (detection limits for apparent opacity).

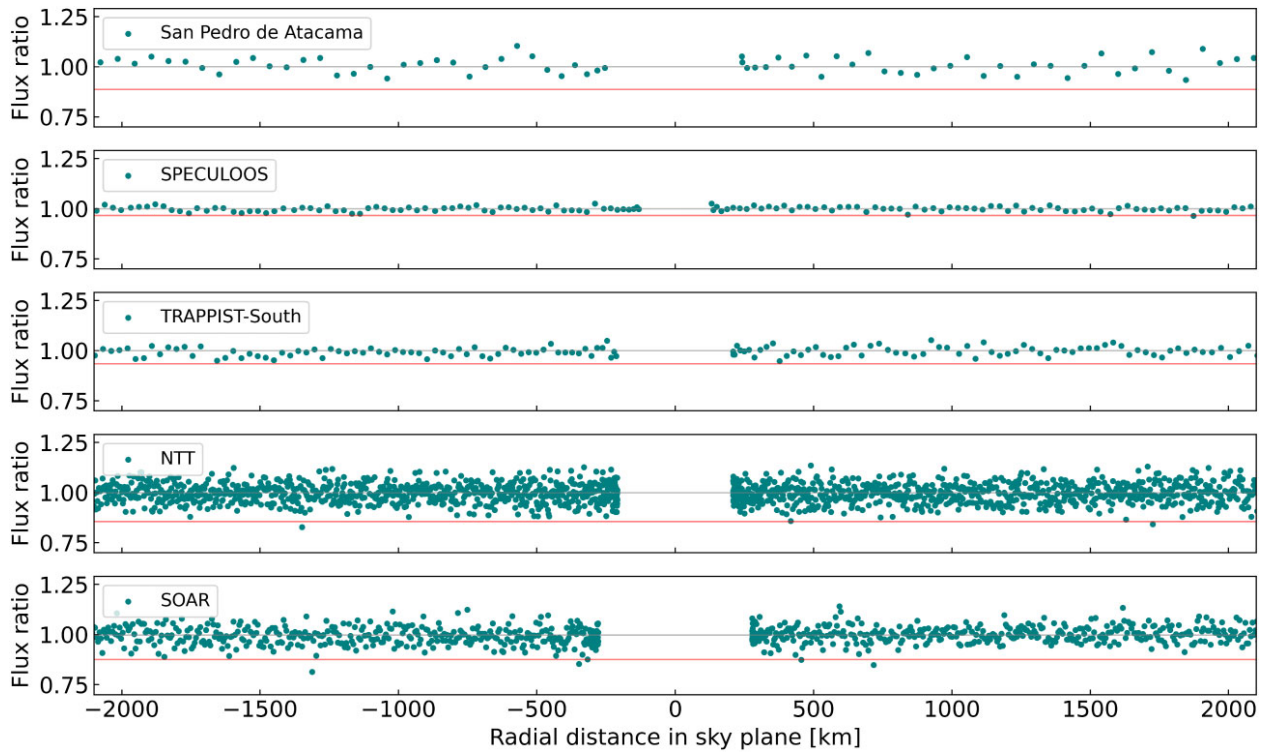

**Figure D1.** Light curves obtained on 2019 Echeclus appulse, with normalized flux plotted against radial distance in the sky plane. The grey horizontal line indicates the mean flux and the red horizontal line presents the depth of an occultation caused by a structure with apparent opacity p′(3σ).

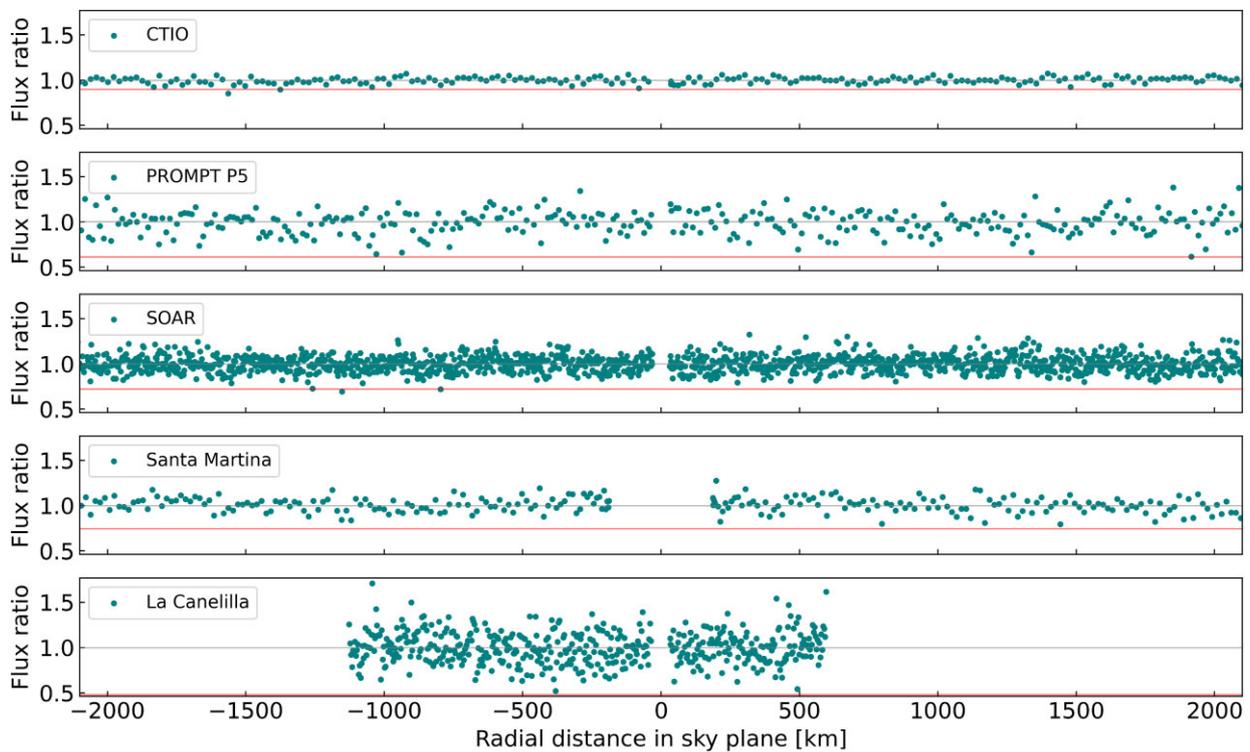

**Figure D2.** Light curves obtained on 2020 occultation event, with normalized flux plotted against radial distance in the sky plane. The grey horizontal line indicates the mean flux, and the red horizontal line presents the depth of an occultation caused by a structure with apparent opacity p$'$(3$\sigma$). The detection of the main body in the SOAR and La Canelilla light curve was omitted in this analysis.

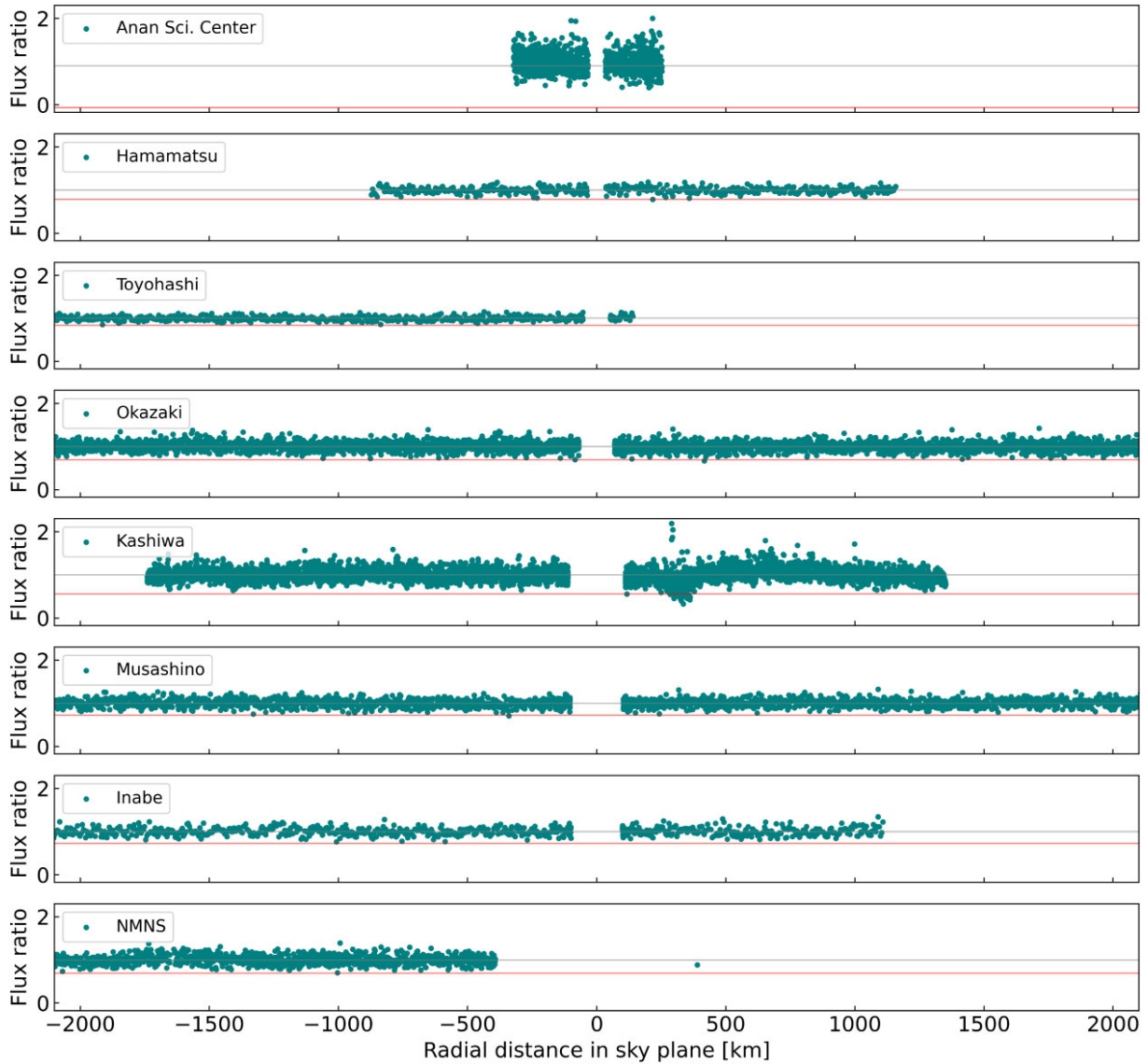

**Figure D3.** Light curves obtained on 2021 stellar occultation, with normalized flux plotted against radial distance in the sky plane. The grey horizontal line indicates the mean flux, and the red horizontal line presents the depth of an occultation caused by a structure with apparent opacity p'(3$\sigma$). The detection of the main body at the Anan Science Center light curve was omitted in this analysis. The Toyohashi acquisition ended just after the closest approach, hence the lack of data in the graph. The same occurred with the National Museum of Nature and Science data, with the acquisition about 400 km outside the expected closest approach position.

## APPENDIX E: MODELED LIGHT CURVES

Positive light curves were modeled with the Fresnel scale, exposure time, star apparent diameter, and square-well model to determine the instants of ingress and egress (Table 3). Fig. E1 presents these modeled light curves in more detail.

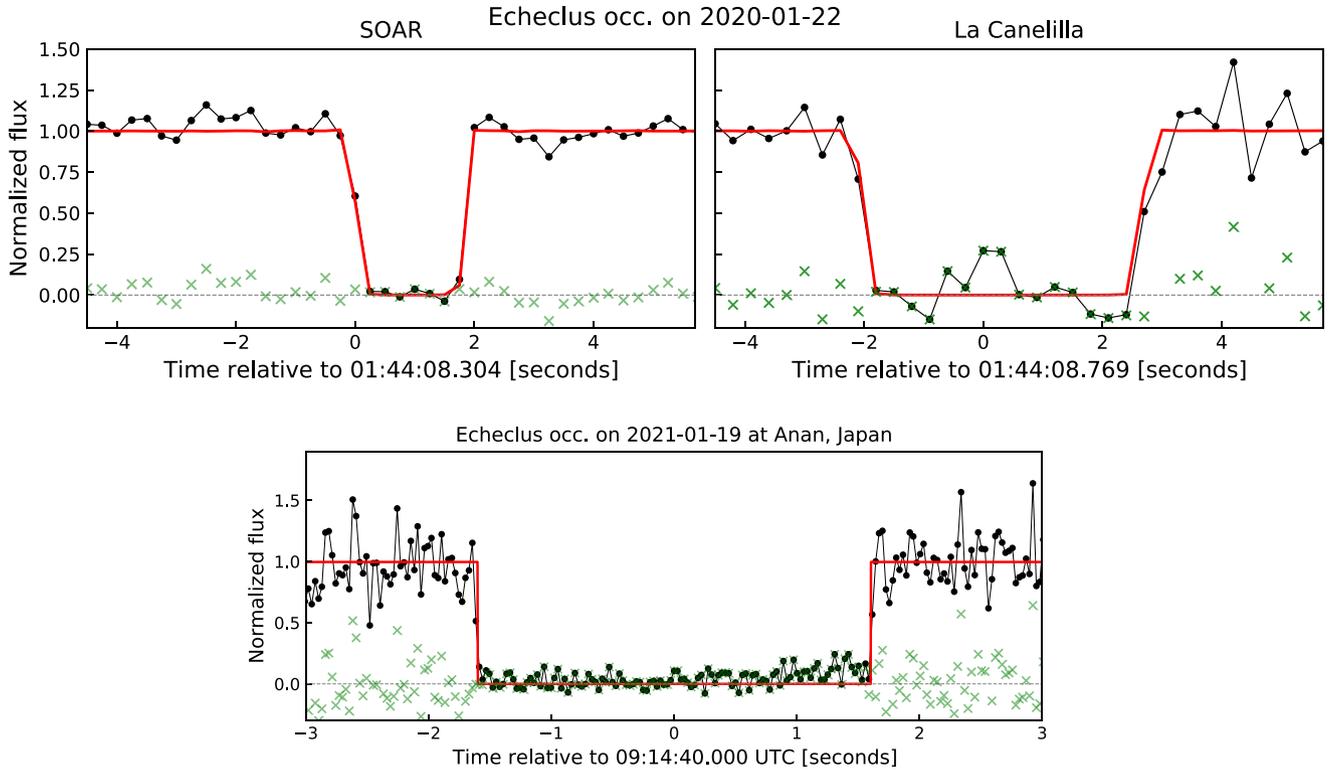

**Figure E1.** Modelled light curves for 2020 (top) and 2021 (bottom) stellar occultations. The black curve represents the data, and the red curve is the modelled light curve. The green markers represent the residuals. The zero instant in graphs is the closest approach in each site.

## APPENDIX F: OCCULTATION MAPS

The Figs F1, F2, and F3 contain the maps of 2019, 2020, and 2021 occultations, showing the shadow's path and the geographical position of all the sites involved in observations.

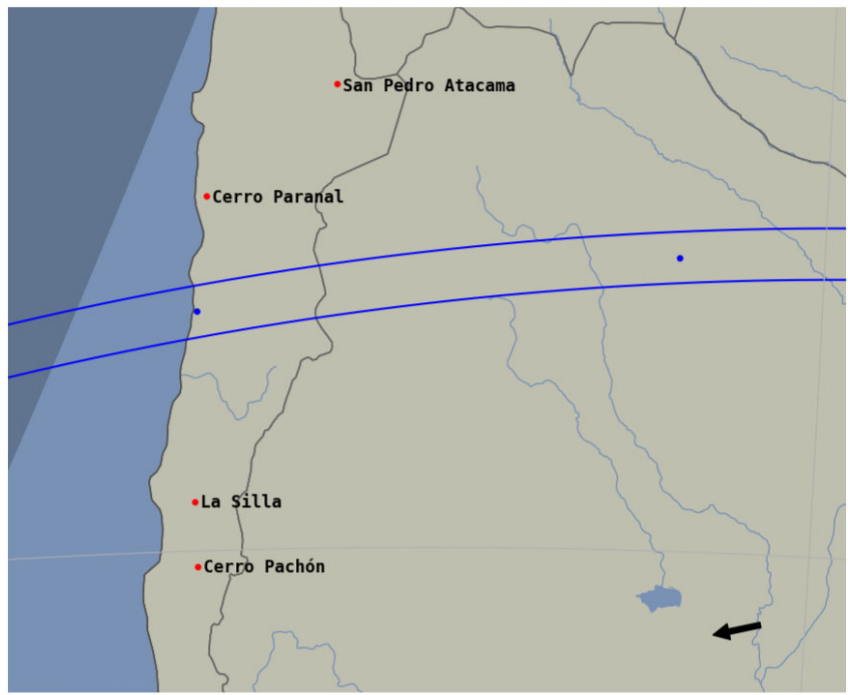

**Figure F1.** Map with the theoretical shadow's path of the 2019 appulse represented by the blue lines. The sites that participated of this observational campaign are indicated by the red markers.

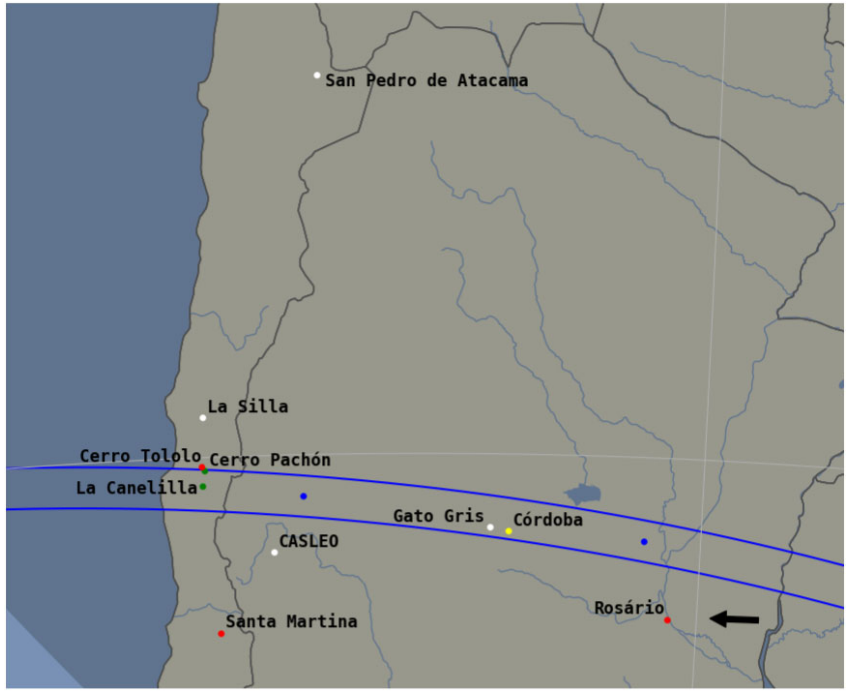

**Figure F2.** Map of the event observed on 2020 January 22. The green dots represent the sites that successfully observe the occultation. The sites which not detect this event are represented in red (observed, but no event was seen), white (overcast), and yellow (technical problems). The blue lines represent the Echeclus shadow path, and the arrow shows the direction of shadow movement.

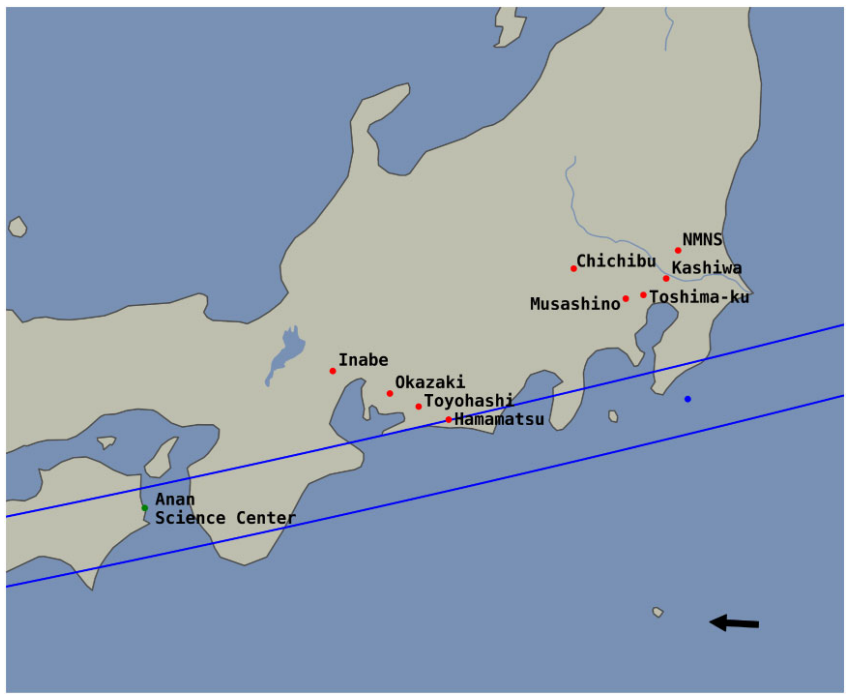

**Figure F3.** Map of the event observed on 2021 January 19. The green dot represents the site that successfully observes the occultation. The sites not detecting this event are represented by red dots (observed, but no event was seen). The blue lines represent the Echeclus shadow path, and the arrow shows the direction of shadow movement.

[1]*Observatório Nacional/MCTI, R. General José Cristino 77, CEP 20921-400 Rio de Janeiro - RJ, Brazil*
[2]*Laboratório Interinstitucional de e-Astronomia - LIneA, Av. Pastor Martin Luther King Jr 126, CEP: 20765-000 Rio de Janeiro, RJ, Brazil*
[3]*Federal University of Technology - Paraná (UTFPR-Curitiba), Rua Sete de Setembro, 3165, CEP 80230-901, Curitiba, PR, Brazil*
[4]*LESIA, Observatoire de Paris, Université PSL, Sorbonne Université, Université de Paris, CNRS, 92190 Meudon, France*

[5]*Institute of Physics, Federal University of Uberlândia, Av. Joao Naves de Avila, CEP 38408-100, Uberlândia-MG, Brazil*

[6]*UNESP - São Paulo State University, Grupo de Dinâmica Orbital e Planetologia, CEP 12516-410, Guaratinguetá, SP, Brazil*

[7]*Instituto de Astrofísica de Andalucía - Consejo Superior de Investigaciones Científicas, Glorieta de la Astronomía S/N, E-18008, Granada, Spain*

[8]*Space Science and Technology Centre, Curtin University, Kent Street, Bentley, Perth, Western Australia 6102, Australia*

[9]*Universidade Federal do Rio de Janeiro - Observatório do Valongo, Ladeira Pedro Antônio 43, CEP 20.080-090, Rio de Janeiro - RJ, Brazil*

[10]*Institut Polytechnique des Sciences Avancées IPSA, 63 boulevard de Brandebourg, 94200 Ivry-sur-Seine, France*

[11]*Institut de Mécanique Céleste et de Calcul des Éphémérides, IMCCE, Observatoire de Paris, PSL Research University, CNRS, Sorbonne Universités, UPMC Univ Paris 06, Univ. Lille, France*

[12]*Universidade Estadual de Ponta Grossa, Av. General Carlos Cavalcanti - Uvaranas CEP 84030-000, O.A. - DEGEO, Ponta Grossa (PR), Brazil*

[13]*Japan Occultation Information Network (JOIN), Japan*

[14]*International Occultation Timing Association - East Asia, 281, Kasadashin-den, Inabe, Mie, 511-0205, Japan*

[15]*STAR Institute, University of Liège, Allée du 6 avril, 19C, B-4000 Liège, Belgium*

[16]*Obs. Ast. Córdoba, Universidad Nacional de Córdoba-CONICET Laprida 854, X5000 BGR, Córdoba, Argentina*

[17]*Observatorio Astronómico El Gato Gris, San Luis CPA X5155DXA, Tanti - Valle de Punilla - Córdoba - Argentina*

[18]*AIM, CEA, CNRS, Université Paris-Saclay, Université de Paris, 91191 Gif-sur-Yvette, France*

[19]*Panamanian Observatory in San Pedro de Atacama - OPSPA, San Pedro de Atacama, Chile*

[20]*Florida Space Institute, UCF, 12354 Research Parkway, Partnership 1 building, Room 211, Orlado, USA*

[21]*Astrobiology Research Unit, Université de Liège, Allée du 6 Août 19C, B-4000 Liège, Belgium*

[22]*National Museum of Nature and Science, 4-1-1 Amakubo, Tsukuba, Ibaraki 305-0005, Japan*

[23]*Complejo Astronómico El Leoncito (CASLEO), Av. España 1512 Sur J5402DSP, San Juan, Argentina*

[24]*San Pedro de Atacama Celestial Explorations - SPACE, Solor, San Pedro de Atacama, Chile, Chile*

[25]*Instituto de Astronomía y Fiísica del Espacio, (CONICET-UBA), Intendente Güiraldes S/N, CABA, C1428ZAA, Argentina*

[26]*Facultad de Ciencias Astronómicas y Geofiísicas, Universidad Nacional de La Plata, Paseo del Bosque S/N, La Plata, B1900FWA, Argentina*

[27]*Instiuto de Tecnologia e Ingeneria, UNAHUR, Vergara 2222, Hurlingham, Prov. de Buenos Aires, Argentina*

[28]*Instituto de Astrofísica, Facultad de Física, Pontificia Universidad Católica de Chile, Av. Vicuña Mackenna 4860, Santiago, Chile*

[29]*Physics and Astronomy Department, Appalachian State University, Boone, NC 28608, USA*

[30]*Estación Astrofísica Bosque Alegre, Córdoba, Argentina*

[31]*School of Physics & Astronomy, University of Birmingham, Edgbaston, Birmingham B15 2TT, UK*

[32]*Department of Electrical Engineering and Center of Astro Engineering, Pontificia Universidad Católica de Chile, Av. Vicuña Mackenna 4860, Santiago, Chile*

This paper has been typeset from a TeX/LaTeX file prepared by the author.